\documentclass[dvips]{arxstspdf}
\usepackage{graphicx}
\usepackage{flushend}
\usepackage{stfloats}

\volume{26}
\issue{4}
\pubyear{2011}
\firstpage{517}
\lastpage{527}
\doi{10.1214/09-STS295}

\begin{document}
\begin{frontmatter}

\title{The 2004 Venezuelan Presidential Recall Referendum: Discrepancies
Between Two~Exit Polls and Official Results}
\runtitle{Exit polls in the Venezuelan recall referendum}

\begin{aug}
\author[a]{\fnms{Raquel} \snm{Prado}\ead[label=e1]{raquel@ams.ucsc.edu}}
and
\author[b]{\fnms{Bruno} \snm{Sans\'{o}}\corref{}\ead[label=e2]{bruno@ams.ucsc.edu}}
\runauthor{R. Prado and B. Sans\'{o}}
\pdfauthor{Raquel Prado, Bruno Sanso}

\affiliation{University of California, Santa Cruz}

\address[a]{Raquel Prado is Associate Professor, Department of Applied Mathematics and Statistics,
Baskin School of Engineering, University of California, Santa
Cruz,
1156 High Street, Mail Stop: SOE,
Santa Cruz, California 95064, USA \printead{e1}.}
\address[b]{Bruno Sans\'{o} is Professor and Department Chair, Department of
Applied Mathematics and Statistics, Baskin School of Engineering, University of California, Santa
Cruz,
1156 High Street, Mail Stop: SOE,
Santa Cruz, California 95064, USA \printead{e2}.}

\end{aug}

\begin{abstract}
We present a simulation-based study in which the results of two major
exit polls conducted during the recall referendum that took place in
Venezuela on August 15, 2004, are compared to the official results of
the Venezuelan National Electoral Council ``Consejo Nacional
Electoral'' (CNE).  The two exit polls considered here were conducted independently
by S\'{u}mate, a nongovernmental organization, and Primero Justicia, a
political party.  We find significant discrepancies between
the exit poll data and the official CNE results in about 60\% of the
voting centers that were sampled in these polls. We show that
discrepancies between exit polls and official results are not due to a
biased selection of the voting centers or to problems related to the
size of the samples taken at each center.  We found discrepancies in all
the states where the polls were conducted.  We do not have enough
information on the exit poll data to determine whether the observed
discrepancies are the consequence of systematic biases in the selection
of the people interviewed by the pollsters around the country. Neither
do we have information to study the possibility of a high number of false or
nonrespondents. We have limited data suggesting that the discrepancies
are not due to a drastic change in the voting patterns that occurred
after the exit polls were conducted. We notice that the two exit polls
were done independently and had few centers in common, yet their overall
results were very similar.
\end{abstract}


\end{frontmatter}

\section{Introduction}

A presidential recall referendum (RR) took place in Venezuela on August
15, 2004. A tense political debate preceded the RR, the main issue
being the timing and the validity of the process that led to the RR
actually taking place. The Organization of American States (OAS) sent
a delegation chaired by its Secretary General to negotiate a solution.
The Carter Center, led by President Jimmy Carter himself, played an
important role in getting the government and the opposition to agree on a
course of action. The Consejo Nacional Electoral (CNE) was the official
body in charge of the organization of the RR.\looseness=-1

Since the RR was seen by all parties involved as a~pivotal
event, several organizations set up schemes to collect exit poll data.
In this work we study data from two exit polls collected independently
by S\'{u}\-mate, a Venezuelan nongovernmental organization (NGO) and
Primero Justicia, a Venezuelan political party. S\'{u}mate
(\url{http://sumate.org/index.html})\break
\mbox{defines} its mission as that of ``building
democracy.'' It played a leading role in the process of collecting the
signatures needed to call the RR. Primero Justicia (\url{http://www.primerojusticia.org.ve})
is a~relatively young
party that campaigned actively to recall the President.  The exit poll
samples
were collected during the day when the RR took place, at a number of
voting centers around the country.
As seen in Table \ref{tab:official}, the official results of
the recall (available from \texttt{\href{http://www.cne.gov.ve}{www.cne.gov.ve}}) were
40.6\% in favor of recalling the president (Yes vote) and
59.1\% against recalling the president (No
vote).  There was a percentage of 0.3\% of invalid votes. These results
are in sharp contrast with the exit poll results. According to S\'{u}mate,
the percentage of Yes votes was 60.7\% and according to Primero
Justicia, the percentage of Yes votes was 60.5\%.

\begin{table}
\caption{Official results of the 2004 Venezuelan recall referendum}\label{tab:official}
\begin{tabular*}{\columnwidth}{@{\extracolsep{\fill}}lcc@{}}
\hline
& \textbf{Number votes} & \textbf{\% of votes}  \\
\hline
Registered voters & 14,037,900 & 100\%\phantom{0.0} \\
Casted votes & \phantom{0}9,815,631 & \phantom{0}69.92\% \\[5pt]
& \textbf{Number votes} & \textbf{\% of casted votes}  \\
\hline
Yes votes & 3,989,008  & 40.64\% \\
No votes & 5,800,629  & 59.10\% \\
Invalid votes & \phantom{0.0}25,994 & \phantom{0}0.26\% \\
\hline
\end{tabular*}
\vspace*{-3pt}
\end{table}

The large observed discrepancies between these exit polls and the
actual official results immediately triggered discussions among experts
and nonexperts in Venezuela questioning the validity of the polls
and the official results.
One of the arguments raised against the exit polls was that
the selection of the polled centers was biased. Another was that exit
polls were conducted only until early in the afternoon, while many of the
voting centers stayed open until late at night. Yet another argument was
that the exit polls were biasedly conducted by interviewers who were
prone to choose pro-Yes voters and that No-voters were less inclined to
respond to these polls.

Hausmann and Rigob\'{o}n (\citeyear{HauRig2004}) present a compre\-hensive
discussion of several issues regarding the possibility of fraud in the
recall referendum. Other related references include Delfino and Salas
(\citeyear{DelSal2009}) and Taylor (\citeyear{Tay2009}). In particular,
for the S\'{u}mate and Primero Justicia exit polls, Hausmann and Rigob\'{o}n
show that there are no significant differences between the
official results for the centers in those polls and the official results
for the overall population, thus indicating that the selection of
the polled centers was not particularly biased.

In this study we find significant discrepancies between the S\'{u}mate and
PJ exit poll results and the official results for the majority of the
voting centers. We also find (via a simulation study) that the
discrepancies are neither due to chance nor to the exit polls taking too
small a sample for each center. While most of the centers in the exit
polls were operated with voting machines, some manual centers were also
sampled. We consider this small subgroup separately, since the analysis
in these cases is complicated by the presence of invalid votes. Invalid
votes are virtually nonexistent for the automatic centers.

S\'{u}mate has produced a report on the entire refe\-rendum process
(S\'{u}mate, \citeyear{Sum2004}), which is available from their web page. The Carter Center has
produ\-ced two reports, one on the audit of the RR \mbox{results} (The Carter
Center, \citeyear{TCC2004}) and a final report about observing the RR process (The
Carter Center, \citeyear{TCC2005}). Both reports are available at
\texttt{\href{http://www.cartercenter.org}{www.cartercenter.org}}.

The exit poll data analyzed here were provided by S\'{u}mate. The data are
available from S\'{u}mate upon request. The official data on the recall
referendum were obtained  from the official CNE web site
\texttt{\href{http://www.cne.gov.ve}{www.cne.\break gov.ve}}.

\section{Exit Poll Data}\vspace*{3pt}

\subsection{S\'{u}mate Exit Poll}

The exit poll conducted by S\'{u}mate consisted of a sample of 269 voting
centers, out of 8279 total centers. These were located in 23 of the 24
states and federal entities. The exception
being the State of Delta Amacuro. Only one center was polled in each of the
states of Cojedes and Amazonas. In all other states at least two centers
were polled. Voting centers in Venezuelan embassies and
consulates were not included in the polling. A total of 23,827 people
were interviewed in these centers. This corresponds to a population of
945,074 voters registered in such polled centers. The exit poll was designed
by S\'{u}mate and the American polling firm Penn, Schoen and Berland
Associates (PSB).  According to S\'{u}mate, the pollsters were volunteers
who were trained and supervised by S\'{u}mate and PSB for more than a
month. They were instructed to follow a protocol to guarantee that the
sample had the least possible bias. In particular, strong emphasis was
given to the fact that the pollsters should not be identified as members
of S\'{u}mate or any other political group. Samples were collected by
asking selected people coming out of the voting center to deposit an
envelop with their voting option in a closed ballot box.  This was
the only question asked, no additional information about the person
interviewed (e.g., age, sex, income, etc.) was recorded and linked to his
or hers voting option.  People were selected to reflect the proportion
of gender and age distribution in the center. Each center had a target
sample size per hour for each gender and for each one of three age
groups. A sample of an hourly reporting form is presented in Table~\ref{sample_sheet}. Pollsters were instructed to avoid interviewing more
than one person from a given cluster of people.  Ballots from all the
voters interviewed were deposited in the same box and results were
reported to a supervisor every two hours.

\begin{table}
\tabcolsep=0pt
\caption{Example of hourly reporting form for the S\'{u}mate exit poll corresponding to the time 7:00--8:00 am}\label{sample_sheet}
\begin{tabular*}{\columnwidth}{@{\extracolsep{4in minus 4in}}lcccccc@{}}
\hline
 & \multicolumn{6}{c@{}}{\textbf{Age}}\\[2pt]
\cline{2-7}
& \multicolumn{2}{c}{$\bolds{<}$\textbf{30}} & \multicolumn{2}{c}{\textbf{30--50}} & \multicolumn{2}{c@{}}{$\bolds{>}$\textbf{50}} \\
\ccline{2-3,4-5,6-7}
\textbf{Gender} & \textbf{Target} & \textbf{Polled} & \textbf{Target} & \textbf{Polled} &
\textbf{Target} & \textbf{Polled} \\[-2pt]
\hline
Male   & 1 & & 2 & & 1 & \\
Female & 2 & & 2 & & 1 & \\
 Total & 3 & & 4 & & 2 & \\
\hline
\end{tabular*}
\end{table}

The sampling scheme was designed to collect data from
the chosen voting centers between 7:00 am and 5:00 pm, since the
Venezuelan electoral law states that voting centers have to be opened at
7:00 am and should close by 4:00 pm, but should remain open as long as
there are voters in line. The voting process was extremely slow due to a
historically large attendance of the voters and the introduction of new
voting technology, such as fingerprint readers and automatic voting
machines. Because of this, the CNE extended the closing hour of the
voting centers twice during the afternoon of the RR day, first
indicating that centers had to remain open until 9 pm and then extending
the closing time until 12:00 pm (see the reports of the Carter Center
and S\'{u}mate for a summary of some of the facts related to the
procedures that need to be followed during election\vadjust{\goodbreak} day and the actual
RR process). For this reason, only 24 out of the 269 centers
chosen to be polled by S\'{u}mate were polled until 11:00 pm.

We have access to a database where the samples are recorded every two
hours, as they were reported by the pollsters. We observed that the data
collection process was, for most centers, very regular. We were able to
calculate the overall target for each center and compare that to the
effectively observed sample. This is important to establish if a center
was strongly above or below target. We notice that being below target
does not provide useful information about the nonresponse. Indeed, there
could be many factors affecting the fact that a pollster collected less
samples than planned.

\setcounter{figure}{1}
\begin{figure*}[b]
\vspace*{-3pt}
\includegraphics{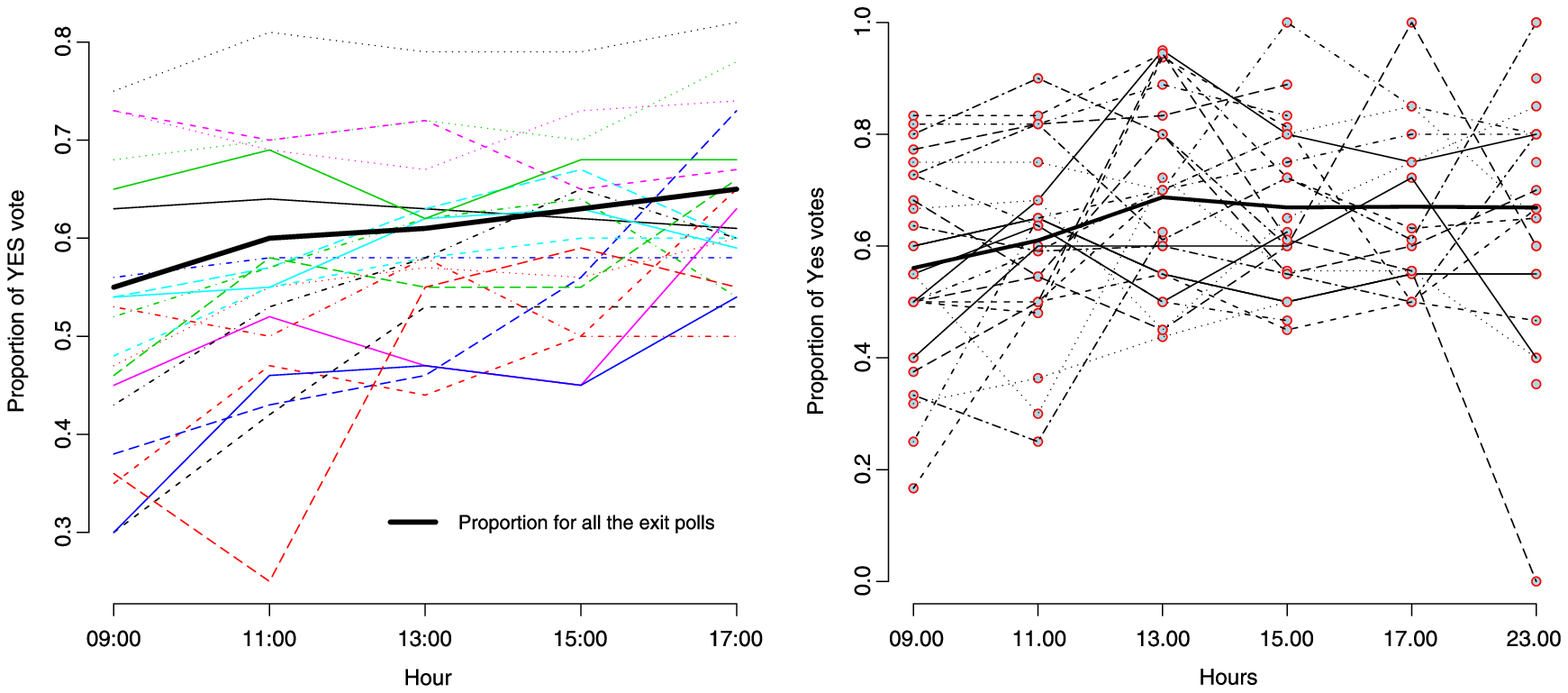}

\caption{Proportion of Yes votes at reporting times for the S\'{u}mate
exit poll: proportions for the states (left panel), proportions for the
centers that were polled until 23:00 (right panel).}\label{fig:perhour}
\end{figure*}

\subsection{Primero Justicia Exit Poll}

The exit poll conducted by Primero Justicia (PJ) consisted of a sample
of 258 voting centers located in 21 of the 24 Venezuelan states and
federal entities. The three states excluded from the sample were
Amazonas, Cojedes and Delta Amacuro. No embassies or consulates were
included in the sample. A total of 12,347 people were interviewed. This
corresponds to a population of 1,151,980 voters registered in the
polled centers.  The protocol that the interviewers followed to collect
the sample was similar to that used by S\'{u}mate. Samples were
collected from 7:00 am to 3:00 pm.  There was a target number of samples
to be obtained per hour. Reports were sent to a central office six times
during the day.  Interviewers worked in pairs and were given specific
instructions on the way to perform the interview so that minimum bias
would result. The PJ poll set targets for the number of interviews that
were smaller than the ones set by S\'{u}mate and sampled only up to 3:00
pm.  This produced a total sample size of roughly a half of that of
S\'{u}mate.  For this exit poll we only have the total number of samples
taken during the day.

\section{Data Analysis}

In this analysis we only consider centers for which more than 20 samples
were collected in either of the two exit polls. In addition, we exclude centers
from the S\'{u}mate exit poll for which the number of people interviewed was
50\% smaller or 20\% higher than the target sample size. We exclude the
centers that were strongly above their targets since this is taken as an
indication that the interviewer was in violation of the protocol. In
particular, there could be many unsolicited answers that could bias the
sample. We were\vadjust{\goodbreak} left with  data for a total of 497 centers. From these
497 centers, 464 were automatic and 33 were manual. The S\'{u}mate
and PJ polls have 27 common centers, all of them automatic.

\setcounter{figure}{0}
\begin{figure}[b]

\includegraphics{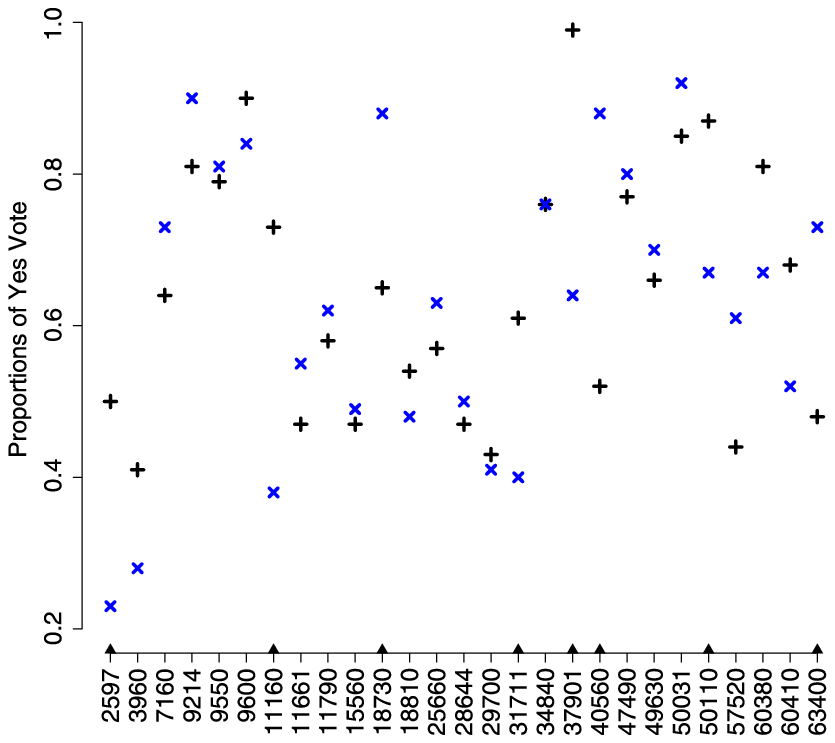}

\caption{Comparison of the common exit polls. The results from
the S\'{u}mate exit poll (indicated by $+$) and the PJ exit poll
(indicated by $\times$) are shown.
Centers in which the two polls differ significantly are highlighted
in the bottom margin. Significance is established at the 1\% level.
Numbers in the $x$ axis correspond to the CNE coding of the centers.}\label{fig:comparison.both}
\vspace*{-6pt}
\end{figure}

A comparison of the results obtained by the two exit polls in the common
centers shows that in 19 out of the 27 cases the exit polls produce
compatible results.\vadjust{\goodbreak} Figure \ref{fig:comparison.both} gives a graphical
comparison of the samples. The significance is established using a~$z$
test at the 1\% level (for details about the $z$ test see, for example,
DeGroot and Schervish, \citeyear{DeGSch2002}). We closely inspected the data
of the 8 centers where the two exit polls differ significantly. We
found that in~2 of these 8 centers there are reasons to believe
that one of the two exit polls might be biased in favor or against the
Yes vote. This may be due to problems related to either a small sample
size or a biased selection of the people interviewed. The
official recall results are compatible with at least one of the two exit
polls, usually the one from S\'{u}mate, in 7 of these 8 centers.

The availability of data reported every two hours for the S\'{u}mate exit
polls provides information about possible trends in the voting pattern
during the day. The left panel in Figure \ref{fig:perhour} shows the
proportion of Yes votes per state at the five reporting times for all
the states that were polled. We see
no obvious pattern for the state data. We do observe a slight increasing
pattern for the overall proportion of Yes votes, as suggested by the
thick black line that corresponds to the mean. The right panel shows the
proportion of Yes votes
for the centers that were sampled until 11 pm. Again, we observe no obvious
pattern or trend here except for a slight increase in the overall
proportion of Yes votes during morning hours.

\setcounter{figure}{2}
\begin{figure*}

\includegraphics{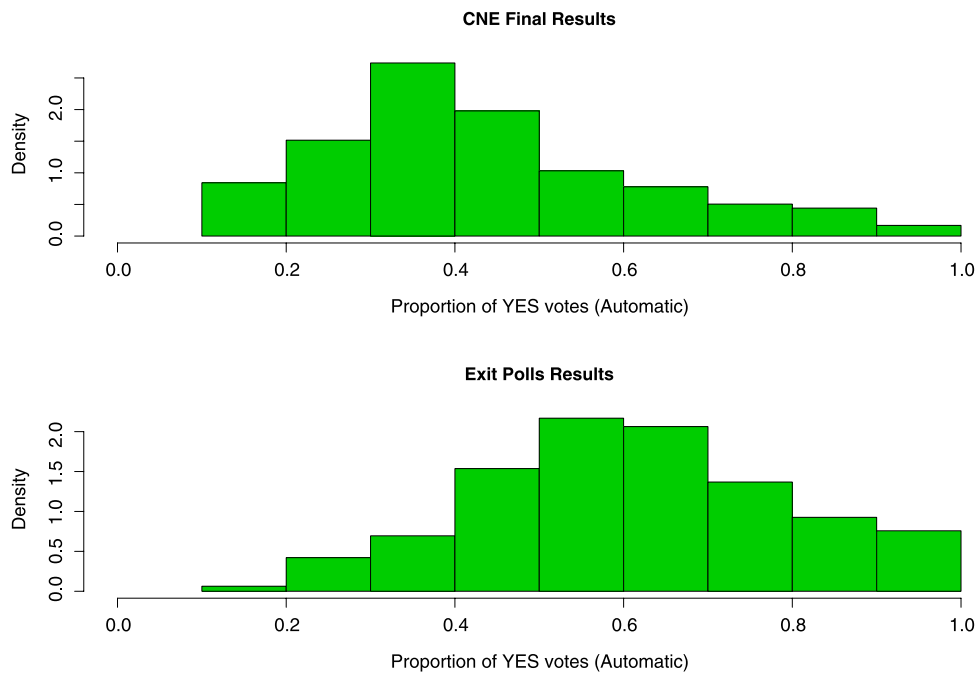}

\caption{Distribution of Yes vote proportions by center: official
results (top panel), exit poll results (bottom panel).}\label{fig:comparison.cne}
\vspace*{6pt}
\end{figure*}

In order to determine if the automatic and manual centers exhibited a
different behavior, we analyzed the data from these two classes of centers
separately.

\subsection{Automatic Centers}\vspace*{3pt}

We begin by comparing the distribution of the proportion of Yes votes obtained in
the exit polls per center and the distribution of the official proportion
of Yes votes obtained in the polled centers.
Figure \ref{fig:comparison.cne}, top panel, displays a histogram of
the proportions of Yes votes reported by the CNE  in the
464 automatic polled centers. The bottom panel of Figure
\ref{fig:comparison.cne} shows a histogram of the proportions
of Yes votes reported by the exit polls in the same 464 automatic centers.
It can be seen from these pictures that the distribution of
the final referendum recall results and the distribution of the exit
poll results differ sharply. Figure \ref{fig:comparison.cne} gives a
clear indication that the differences between the results of the exit
polls and the official ones are not due to a biased choice of the centers.
In fact, for the same centers, we obtain two completely different
distributions of Yes votes.\looseness=1

In order to obtain a more specific quantification of the differences between
the results for a given center, we calculated the likelihood of observing the
samples obtained by the exit polls for such center as follows. Let $y_j$ be the
number of Yes votes observed by the pollster for a given center $j$ and
let~$t_j$ be the size of the sample collected at center $j$ by the
pollster. So, for example, for center $j=1$ (with CNE ID number 400),
located in the Capital District, Municipio Libertador, we have that
$y_{1}=68$ and $t_1=120$. The proportion of Yes votes in the exit
poll for this center is then $y_1/t_1=0.57$. The actual proportion for
this center as reported by the CNE is
$P_{Y,1}=992/2270 = 0.44$. How likely is it that a sample of 68 Yes votes would
be observed in a sample of 120 voters where each voter has probability
44\% of voting yes?

\begin{figure}

\includegraphics{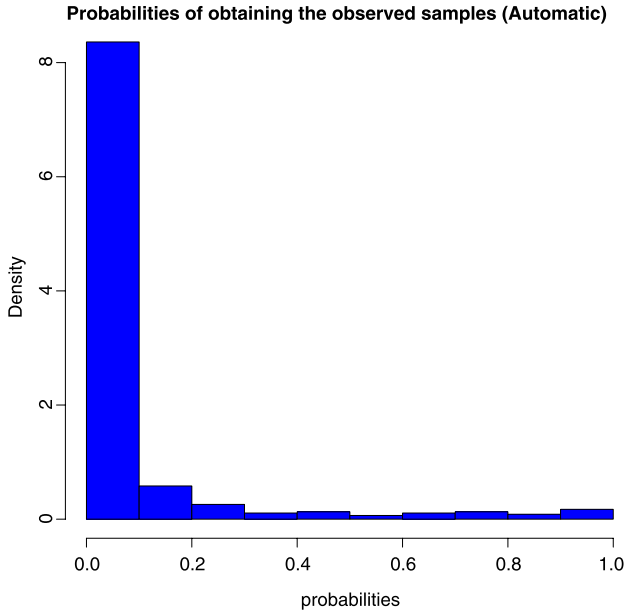}

\caption{Histogram of the probabilities of obtaining the samples
observed in the exit polls using the official results as the
probabilities of a Yes vote.}\label{fig:hist.prob}
\end{figure}

We answer the question by assuming that $y_j$ is a~random sample from a
binomial distribution. Suppose that the true probability of a Yes vote
for center~$j$ is equal to the official proportion of Yes votes for such
center, say, $P_{Y,j}$.  Then
\begin{equation}\label{bin}
y_j \sim \operatorname{Bin}(t_j,P_{Y,j}).
\end{equation}
Using (\ref{bin}), we obtain that the required probability is
$\mathit{Pr}(y_1=68)=0.0015$. Clearly, this probability could be small just
because $y_1$ could take 121 possible values. So we compute $\mathit{Pr}(y_1\geq
68) =0.0035$. We then repeat this calculation for all the
automatic centers. Figure \ref{fig:hist.prob} displays
a histogram of the probabilities for all the centers.
We observe that
about 70\% of these probabilities lie in the interval $(0,0.013)$. That is,
about 70\% of the polled centers' results have a~chance smaller than 1.3\%
of being obtained from a~population where the proportions of Yes votes were
the official proportions.

\subsection{Further Analyses}

\begin{figure*}[t]

\includegraphics{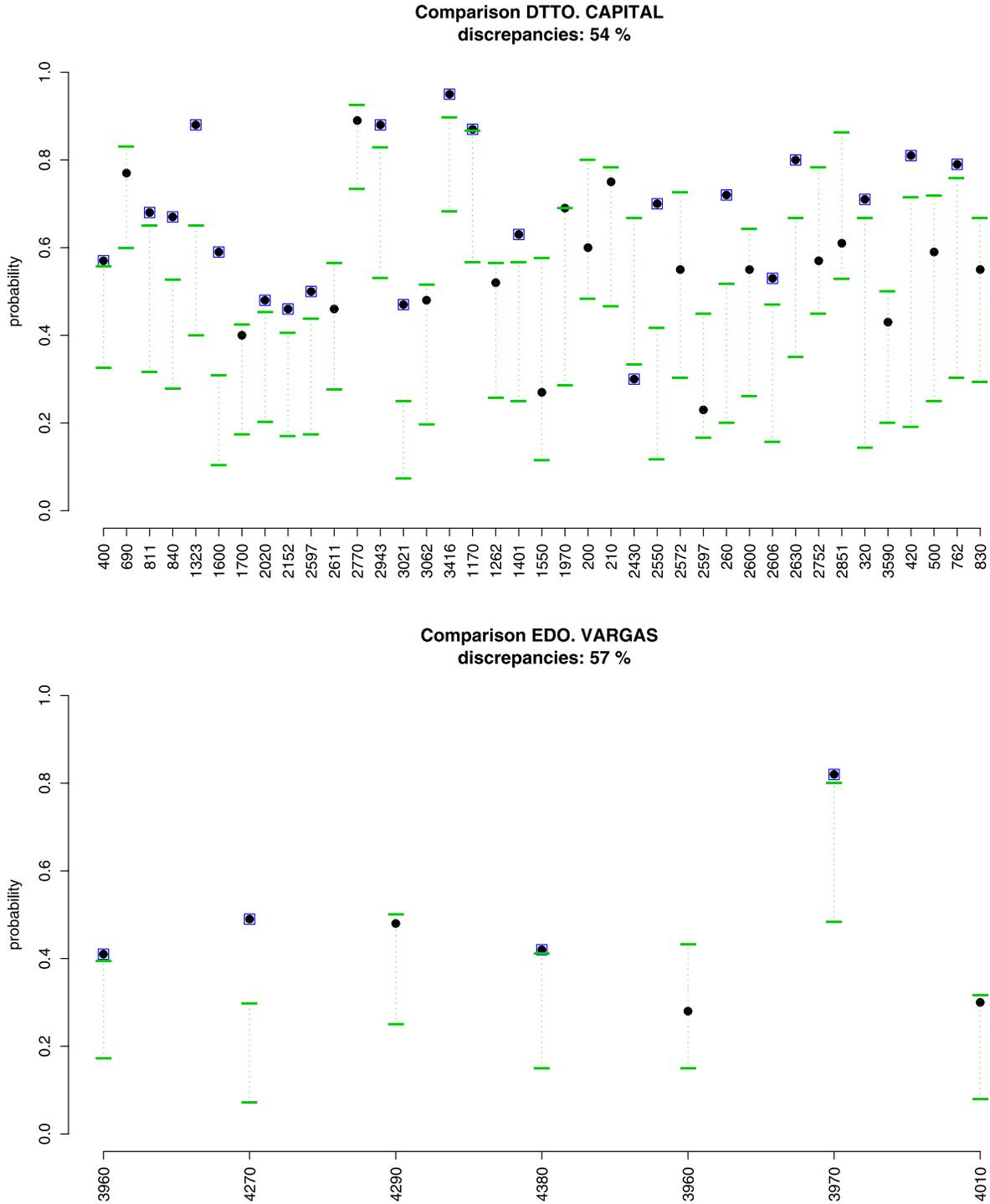}
\vspace*{-4pt}
\caption{99\% probability intervals for the proportions of Yes
votes computed using  5000
simulated samples for each automatic center. The simulated samples were
of the same size as the samples in the exit polls. The probability of a
Yes vote was taken as the official one. The dotted lines indicate the
intervals. The exit poll results are marked with squares when they fall
outside the corresponding intervals.}\label{fig:states1}
\vspace*{-6pt}
\end{figure*}

\begin{figure*}[t]

\includegraphics{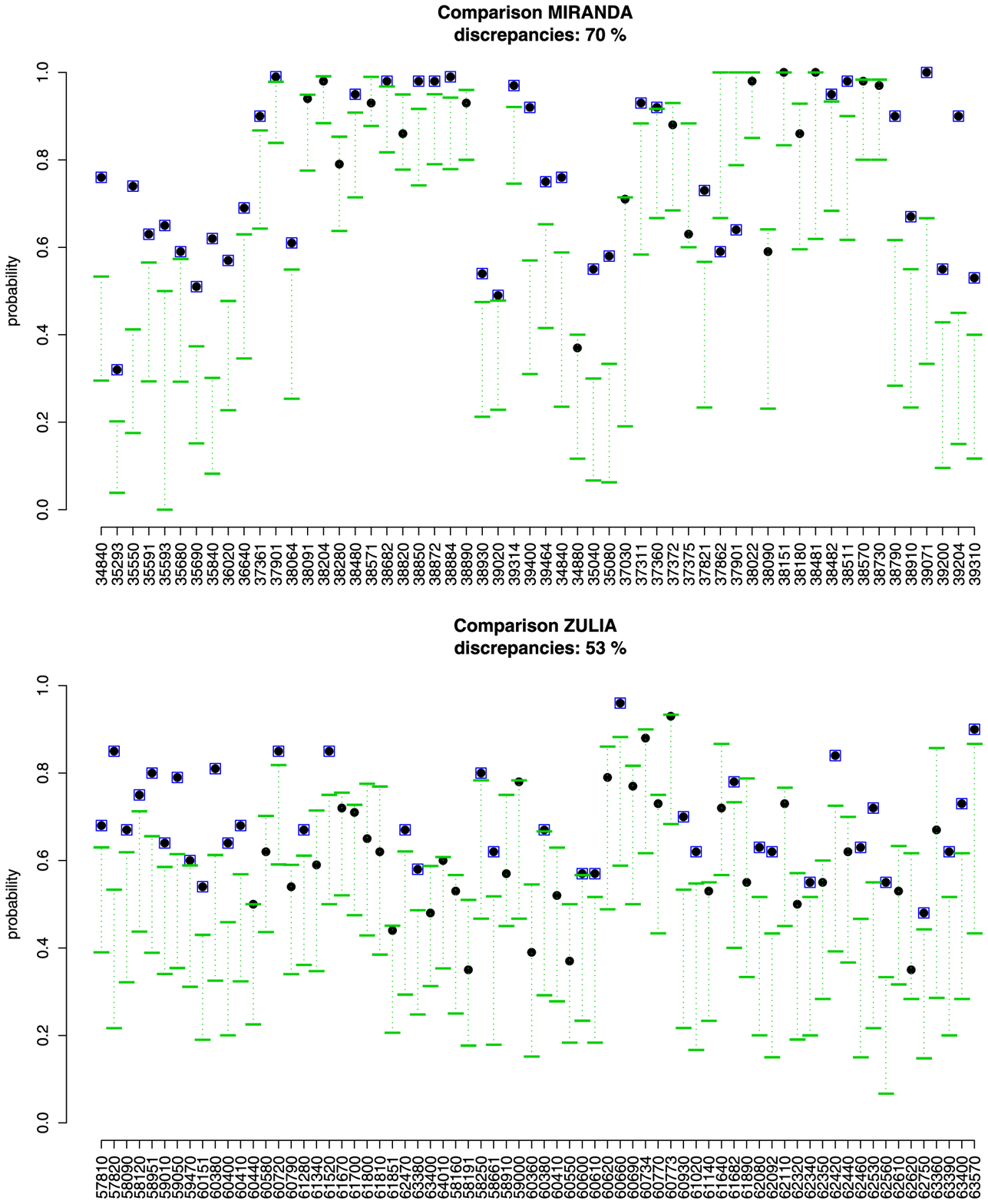}
\vspace*{-4pt}
\caption{99\% intervals for the proportions of Yes votes computed
using 5000 simulated
samples for each automatic center. The simulated samples were of the
same size as the samples in the exit polls. The probability of a Yes
vote was taken as the official one. The dotted lines denote the
intervals.
The exit poll results are marked with squares when they fall outside the
corresponding intervals.}\label{fig:states2}
\vspace*{-6pt}
\end{figure*}

To obtain a better idea of how different the official and the exit polls
results are, we performed a simulation study. Specifically, we simulated
5000 samples of the same size as those of the exit polls, for each
center. In these simulations we set the probability of a Yes vote for
each center at the official proportion of a Yes vote at that center,
denoted $P_{Y,j}$.  In other words, we simulated 5000 samples from
(\ref{bin}), for $j=1,\ldots,464$.  We then computed the 0.05 and 99.5
quantiles (to produce a 99\% interval) of the proportions of Yes votes
in the 5000 samples, and compared the proportions observed in the exit
polls to such intervals. A graphical representation of the results can
be seen in Figures \ref{fig:states1} and \ref{fig:states2} for 4 of the
21 polled states with more than one center. In these figures, a
center where the exit poll proportion falls outside the 99\% interval
is marked with a square. Such a case is labeled as a discrepancy. We
report the percentage of those per state. As an example, in the State of
Miranda the intervals did not cover the exit poll results in 70\% of the
centers.

Zulia, Miranda and the Capital District are the three most
populated states in Venezuela, with approximately 32\% of the total
population. Vargas is a~comparatively small state that is considered as
a~stronghold of the government.  We observe that in about 60\% of the
centers the exit poll result falls above the upper\vadjust{\goodbreak} limit of the
interval.  This happens even when the exit poll predicted that the No
vote would win in a given State, as is the case of Vargas (see bottom
panel in Figure \ref{fig:states1}).  In addition, this is not a peculiar behavior
observed only in certain regions in the country, as can be seen from the
results in Table \ref{tab:discrep}. We observe substantial variability
in both the location and the width of the intervals for centers in some of
the states. Differences in width are not surprising, as exit poll sample
sizes were not uniform. For example, centers 420 and 690, both in Dtto.
Capital, had 21 and 100 polls collected, respectively. The number of
registered voters was 2909 for center 420 and 5021 for center 690.
Additionally, urban regions in Venezuela can have pockets of relatively
affluent areas surrounded by very low income areas. So we can expect
very different voting patterns even in centers that are located nearby. In
other words, both $t_j$ and $P_{Y,j}$ may vary substantially within a
given state. These issues partly explain the disparate distribution of
the intervals in the figures.

\begin{table*}
\caption{Percentage of centers in the exit polls, for states with
more than one polled center, that have significant discrepancies with
the official results}\label{tab:discrep}
\begin{tabular*}{\textwidth}{@{\extracolsep{\fill}}lc@{\hspace*{70pt}}lc@{\hspace*{70pt}}lc@{}}
\hline
\textbf{State} & \textbf{Discrepancies} & \textbf{State} & \textbf{Discrepancies} & \textbf{State} & \textbf{Discrepancies} \\
\hline
Capital & 54\% & Anzo\'{a}tegui & 71\% & Apure & 100\% \\
Aragua & 66\% & Barinas & 67\% & Bol\'{\i}var & 52\% \\
Carabobo & 58\%& Falc\'{o}n & 25\% & Gu\'{a}rico & 64\% \\
Lara& 60\% & M\'{e}rida & 36\% & Miranda& 70\% \\
Monagas & 53\% & Nva. Esparta & 80\%  & Portuguesa & 62\% \\
Sucre & 67\% & T\'{a}chira & 68\% & Trujillo & 54\% \\
Vargas & 57\% & Yaracuy & 70\% & Zulia & 53\% \\
\hline
\end{tabular*}
\end{table*}

\begin{table*}[b]
\caption{Exit poll and official results for the manual centers in
which significant discrepancies were found. The columns correspond to
the Center ID, the number of Yes and No votes in the exit poll, the
proportion of Yes votes in the exit poll $p_{Y,i}$, the number of
official No, Yes and invalid votes (I), the proportion of Yes votes in
the official results $P_{Y,i}$, the closing time (CT) of the exit poll
at the center \mbox{and the closing time of the voting center}}\label{tab:exit}
\begin{tabular*}{\textwidth}{@{\extracolsep{\fill}}lccccccccc@{}}
\hline
\textbf{ID} &  \textbf{Yes} & \textbf{No} & $\bolds{p_{Y,j}}$ & \textbf{Yes} & \textbf{No} & \textbf{I} & $\bolds{P_{Y,j}}$ & \textbf{CT}  &  \textbf{CT} \\
& \textbf{exit poll} & \textbf{exit poll} & & \textbf{CNE} & \textbf{CNE} & & & \textbf{exit poll} &\\
\hline
\phantom{0,}5450  & 63 & 11  & 0.85 &  \phantom{0}77   & \phantom{0}117   & \phantom{0}2  & 0.39 & 3:00 pm & $(*)$ 6:30 pm \\
13,160 & 49 & 35  & 0.58 &  268  &\phantom{0}466    &  66 & 0.34 & 3:00 pm & 8:00 pm \\
14,041 & 44 & 36  & 0.55 &  160   &\phantom{0}820   &    10   &  0.16 & 3:00 pm & NA \\
16,870 & 42 & 46  & 0.48 &  \phantom{0}65   &\phantom{0}265    &    12   &  0.19 & 5:00 pm & $(*)$ 5:10 pm \\
17,480 & 56 & 42  & 0.57 &  183   &\phantom{0}255   &     \phantom{0}0   &  0.42 & 5:00 pm & 9:30 pm \\
21,311 & 50 & 44  & 0.53 &  194   &\phantom{0}425   &    42   &  0.29 & 5:00 pm & 8:30 pm \\
21,630 & 55 & 29  & 0.65 &  189   &\phantom{0}199   &    10   & 0.47 & 3:00 pm & 8:30 pm \\
31,723 & 53 & 27  & 0.66 &  217   &\phantom{0}359   &    18   &  0.37 & 3:00 pm & 9:00 pm \\
34,610 & 67 & 35  & 0.66 &  361   &\phantom{0}538   &    28   &  0.39 & 5:00 pm & 9:07 pm \\
43,140 & 37 & 21  & 0.64 &  \phantom{0}45   &\phantom{0}152    &     \phantom{0}0   &  0.23 & 3:00 pm & $(*)$ NA\\
44,200 & 36&  58  & 0.38 &  \phantom{0}54   &\phantom{0}264    &     \phantom{0}8   &  0.17 & 5:00 pm & $(*)$ NA\\
47,550 & 63 & 27  & 0.70 &  346   &\phantom{0}668   &     \phantom{0}0   &  0.34 & 3:00 pm & 2:00 am\\
48,490 & 58 & 23  & 0.72 &  130   &\phantom{0}277   &     \phantom{0}2   &  0.32 & 3:00 pm & 7:41 pm\\
60,290 & 65 & 29  & 0.69 &  345   &\phantom{0}746   &    44   &  0.30 & 5:00 pm & 7:00 pm\\
11,895 & 34 & 26  & 0.57 & 155   &1678   &    40   &  0.08 & 3:00 pm & NA\\
15,680 & 39 & 21  & 0.65 &  265   &\phantom{0}410   &    30   &  0.38 & 3:00 pm & NA\\
29,330 & 15 & 15  & 0.50 &  \phantom{0}42   &\phantom{0}772    &    48   &  0.05 & 3:00 pm & NA\\
42,808 & 28 & 14  & 0.67 &  231   &\phantom{0}464   &    52   &  0.31 & 3:00 pm & NA\\
\hline
\end{tabular*}
\end{table*}

\subsection{Manual Centers}

We performed a separate analysis of the data corresponding to the manual
centers for two reasons. The first one is that the manual data have
invalid votes. These are almost non existent in the automatic centers.
This implies that the variable corresponding to a vote in a manual center
has three possible outcomes. The second reason is that manual centers
are peculiar. They usually correspond to remote locations and they have
a much smaller number of voters than the automatic ones. Table
\ref{tab:exit} shows the detailed numbers for some of the manual
centers. We can see that most of them had only a~few hundred voters.
This is typical for the 33 manual centers that were included in the exit
poll samples.

\begin{figure*}

\includegraphics{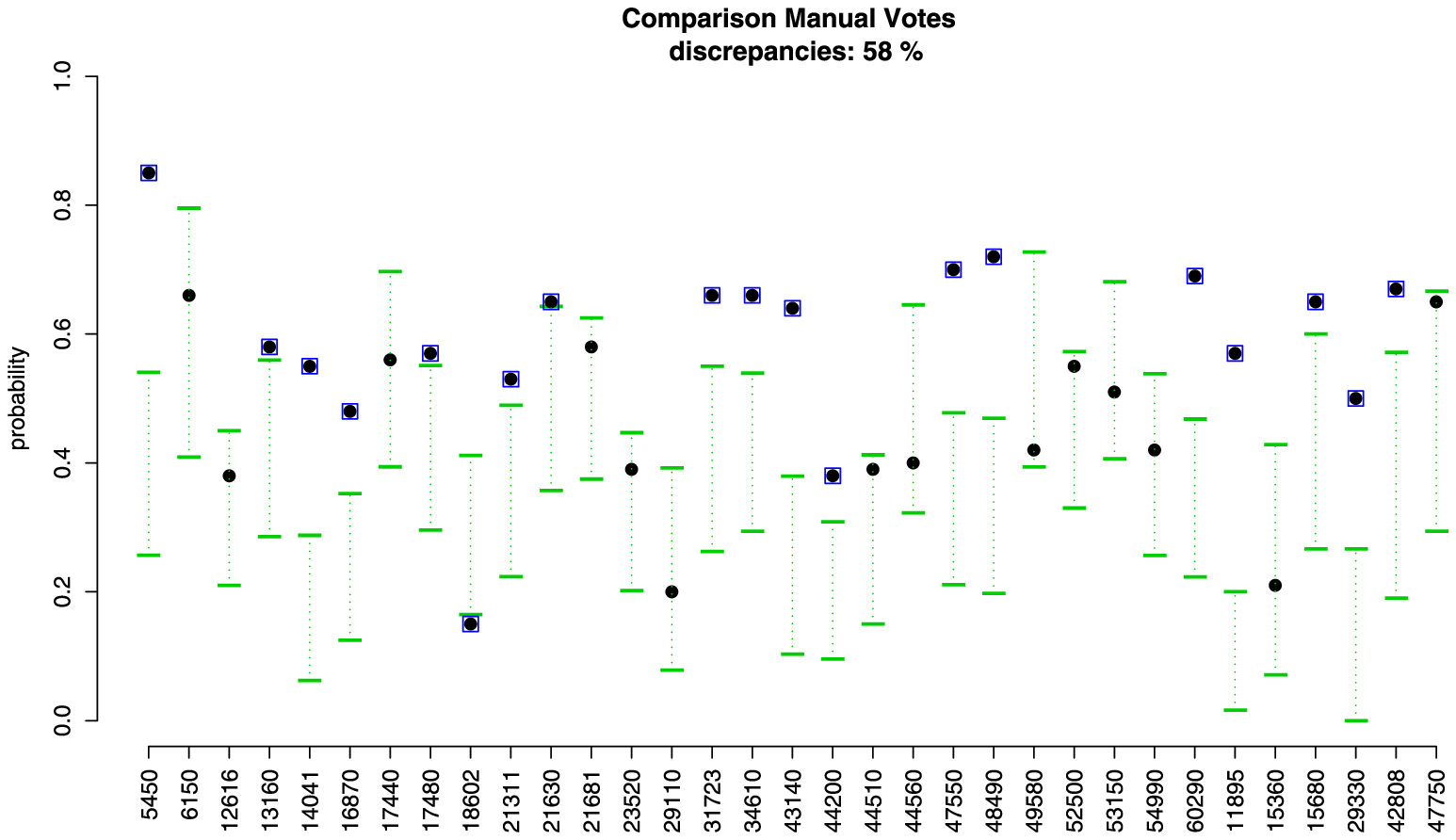}

\caption{99\% probability intervals for the proportions of Yes
votes computed using 5000
simulated samples for each manual center. The simulated samples were of
the same size as the samples in the exit polls. The probability of a Yes
vote for a given center is taken as the official proportion of a Yes
vote in such center. The dotted lines denote the intervals.
Exit poll results are marked with squares when they fall outside the
intervals.}\label{fig:manual}
\end{figure*}

The percentages of invalid votes in the 33 manual centers
considered here go from 0\% to 13.5\%, with an average of 3.5\%.  In
order to take the invalid votes into account, we do the following
calculation.  We assume that the number of Yes, No and invalid votes in
a sample of size $t_j$ taken from center $j$, where the proportions of
Yes votes, No votes and invalid votes are, $P_{Y,j}, P_{N,j}$ and
$P_{I,j}$, respectively, follow a~multinomial distribution. This is
\begin{equation}\label{mult}
(y_{j},n_{j},i_{j}) \sim \operatorname{Mult}(t_j;(P_{Y,j},P_{N,j},P_{I,j})).
\end{equation}
We assume that the proportions of the Yes, No and invalid votes,
$P_{Y,j},P_{N,j}$ and $P_{I,j}$, are the actual
proportions obtained in the recall referendum for each center $j$.
We also assume that the sample size $t_j$ is the same as that taken by the pollsters at
the center.\vadjust{\goodbreak}

Now, for each center $j$, we generate 5000 samples from (\ref{mult}).
We then take the proportion of Yes votes for each one of the 5000
samples as $(y_{j}+i_{j})/t_j$, and use these\vadjust{\goodbreak} values to compute the 99\%
intervals that will be compared against the proportion of Yes votes
observed by the pollsters.  In doing this, we account for the fact that
the pollsters could have interviewed people whose votes resulted
in being invalid.  Note that we are assuming an extreme situation here in which
all the invalid vote samples are actually counted as Yes votes.
Figure~\ref{fig:manual} shows a graphical representation of the
results.\vadjust{\goodbreak}
We find that the exit poll results of 19 out of the
33 manual centers considered present significant discrepancies with the
official results.  In one of these 19 centers the discrepancy occurred as
a result of overestimating the Yes vote due to all the invalid votes
in the simulation being counted as Yes
votes.

Table \ref{tab:exit} gives some interesting insight on the data for the
manual centers where we found large discrepancies. In general, we observe
that the sizes of the exit poll samples are fairly large, relative to the
number of voters. So, assuming that the official results correspond to
the true probabilities of Yes votes, obtaining such large differences
is only possible if the exit poll samples were extremely biased. We
marked with $(*)$ four centers that we found particularly intriguing.
For centers 5450, 16,870, 43,140 and 44,200, the exit polls collected
samples of 38\%, 26\%, 30\% and 23\% of all voters. Of the total Yes votes,
82\%, 65\%, 82\% and 67\% ended up in the exit poll samples, respectively.
If the very low proportions of Yes votes officially reported are correct,
the interviewers did not follow the protocol and were able to
bias the sample.\vspace*{-1pt}

\section{Discussion}\vspace*{-1pt}

The previous analyses involve no sophisticated statistical modeling.
They are based on the assumption that the official CNE results are true.
They provide an exploration of the likelihood that samples like those
in\vadjust{\goodbreak}
the exit polls would be obtained under such assumption. The conclusion
is that, for a~large number of centers, observing samples like the ones
in the exit polls is a very unlikely event, given the official CNE results.
So, clearly, the differences between the predictions of the exit polls
and the actual results at the national level are not due to a~bias in
the selection of the centers. There are significant differences between
the official results and the exit polls in about 60\% of the centers
that are not due to chance. Centers where differences are present are
located in all states, so there does not seem to be any clear geographical
bias in effect in the exit polls.

Clearly, the differences between the exit polls and the official results
could be due to a strong bias in favor of the Yes votes. Such bias would
be the effect of the way samples were collected. To explain the
differences between exit polls and official results, the bias should be
present not just in a few centers, but in about 60\% of them and be
geographically consistent. That is, several hundred pollsters across the
country should have obtained systematically biased samples that favored
the Yes vote, and this in spite of having been precisely instructed to
follow a~protocol to avoid such bias. Given the
protocol, no information about the voters was recorded. So we have
no\vadjust{\goodbreak}
way to associate covariates like, say, gender, age, race, religion or social
status, to voting patterns in a way that would reveal systematic biases
in the selection of the sample.

Unfortunately we have no good estimation of the nonresponse. An
explanation of the discrepancies between the official results and the
exit polls could be that the No voters were less willing to answer than
the Yes voters. Now, suppose that the estimation of Yes vote was about 60\%
when the true value was about 40\%, as in the official result. This
requires that about 33\% of the people interviewed did not respond and
had actually voted No. Again, this phenomenon should have happened all
over the country.

Similar arguments could be given for the false responses. In this case,
about 33\% of the Yes samples should have corresponded to actual No
voters. Such a high level of false responses should have happened even
though the exit polls refer to just one question with a binary secret
answer. Also, the question is about a vote that has already been casted,
so the respondent has no doubts.  The subjects are easily identifiable
and the process of obtaining the sample is quick and simple.

Another explanation for the discrepancies of between exit polls and
official results could be that there were massive numbers of No votes
late in the evening. We have only very limited data shown in Figure
\ref{fig:perhour} to study this possibility. These data do not support
such an explanation. On the contrary, the national average of Yes votes was
slightly increasing over the ten-hour period for which most exit polls
were conducted. We notice that at the time the exit polls were finalized
the proportion of Yes votes was about 60\%. To lower this percentage to
40\% by the end of the evening, the Yes vote proportion should have
plunged to 20\% after 5 pm, for the same number of votes as those casted
during he first ten hours of the voting day.

After the RR, two audits of the automated counts were done by the CNE.
According to the S\'{u}mate report, the first one, known as the ``hot''
audit, was carried out at the time of the closing of the centers in only 84 of 199
preselected centers. Such an audit is also mentioned in the Carter Center's
report. We were unable to find any data related to the result of the hot
audit. The second audit took place three days after the RR and the
opposition parties declined to participate, claiming it was flawed. Two
hundred centers were sampled and some of the ballot boxes from 150 of
those centers audited. Unfortunately, the information regarding which
centers were effectively audited has not been made available by either
the CNE or the Carter Center, who witnessed the audit. Of the 200
centers in the original list, 15 were among those polled by S\'{u}mate and
14, different ones, among those polled by PJ. Unfortunately it has not
been possible to establish if any of these 29 centers were among the
list of 150 audited ones.

The exit polls analyzed in this paper are not the only ones that were
conducted for the RR. We obtained the data of an exit poll conducted by
Proyecto Venezuela, another political party that campaigned actively to
recall the President, based on more than
200,000 voters sampled between 6:30 am and 2 pm. The results are in line
with those of the two exit polls considered in this work. We decided not
to include an analysis of those data here due to the fact that we were unable
to find a good description of the protocol followed by the pollsters.
The Carter Center report mentions an exit poll conducted by the American
firm Evans/McDonough that had the No vote winning with 55\% (see also
Collier, \citeyear{Col2004}). We
were unable to find information about this poll. The web page of the
company has a link to a document containing results from a polling
previous to the RR
(\texttt{%
\href{http://www.evansmcdonough.com/venezuela/VenezuelanPollPre.pdf}%
{www.evansmcdonough.com/venezuela/}
\href{http://www.evansmcdonough.com/venezuela/VenezuelanPollPre.pdf}%
{VenezuelanPollPre.pdf}})
but no mention of an exit poll.

We emphasize that this study does not provide conclusive evidence that
there was fraud in the Vene\-zuelan Presidential recall referendum.
It shows that there are important discrepancies between
the official results and the data obtained on the field during referendum
day.

\end{document}